\documentclass[12pt,preprint]{emulateapj}
\usepackage{graphicx}

\shorttitle{Cyclic Transit Probabilities}
\shortauthors{Stephen R. Kane et al.}
\slugcomment{Submitted for publication in the Astrophysical Journal}

\begin{document}

\title{Cyclic Transit Probabilities of Long-Period Eccentric Planets
  Due to Periastron Precession}
\author{
  Stephen R. Kane\altaffilmark{1},
  Jonathan Horner\altaffilmark{2},
  Kaspar von Braun\altaffilmark{1}
}
\email{skane@ipac.caltech.edu}
\altaffiltext{1}{NASA Exoplanet Science Institute, Caltech, MS 100-22,
  770 South Wilson Avenue, Pasadena, CA 91125}
\altaffiltext{2}{Department of Astrophysics \& Optics, School of
  Physics, University of New South Wales, Sydney, 2052, Australia}


\begin{abstract}

The observed properties of transiting exoplanets are an exceptionally
rich source of information that allows us to understand and
characterize their physical properties. Unfortunately, only a
relatively small fraction of the known exoplanets discovered using the
radial velocity technique are known to transit their host, due to the
stringent orbital geometry requirements. For each target, the transit
probability and predicted transit time can be calculated to great
accuracy with refinement of the orbital parameters. However, the
transit probability of short period and eccentric orbits can have a
reasonable time dependence due to the effects of apsidal and nodal
precession, thus altering their transit potential and predicted
transit time. Here we investigate the magnitude of these precession
effects on transit probabilities and apply this to the known radial
velocity exoplanets. We assess the refinement of orbital parameters as
a path to measuring these precessions and cyclic transit
probabilities.

\end{abstract}



\section{Introduction}

The realization that we have crossed a technology threshold that
allows transiting planets to be detected sparked a flurry of activity
in this direction after the historic detection of HD~209458~b's
transits \citep{cha00,hen00}. This has resulted in an enormous
expansion of exoplanetary science such that we can now explore the
mass-radius relationship \citep{bur07a,for07,sea07} and atmospheres
\citep{ago10,dem07,knu09a,knu09b} of planets outside of our Solar
System. Most of the known transiting planets were discovered using the
transit method, but some were later found to transit after first being
detected using the radial velocity technique. Two notable examples are
HD~17156~b \citep{bar07} and HD~80606~b \citep{lau09}, both of which
are in particularly eccentric orbits. Other radial velocity planets
are being followed up at predicted transit times \citep{kan09a} by the
Transit Ephemeris Refinement and Monitoring Survey (TERMS).

Planets in eccentric orbits are particularly interesting because of
their enhanced transit probabilities \citep{kan08,kan09b}. This
orbital eccentricity also makes those planets prone to orbital
precession. In celestial mechanics, there are several kinds of
precession which can affect the orbital properties, spin rotation, and
equatorial plane of a planet. These have been studied in detail in
reference to known transiting planets, particularly in the context of
the precession effects on transit times and duration
\citep{car10,dam11,hey07,jor08,mir02,pal08,rag09}. One consequence of
these precession effects is that a planet that exhibits visible
transits now may not do so at a different epoch and vice versa.

Here we present a study of some precession effects on known
exoplanets. The aspect which sets this apart from previous studies is
that we are primarily interested in planets not currently known to
transit, particularly long-period eccentric planets which have
enhanced transit probabilities and larger precession effects. We
investigate the subsequent rate of change of the transit probability
to show how they drift in and out of a transiting orientation. We
calculate the timescales and rates of change for the precession and
subsequent transit probabilities and discuss implications for the
timescales on which radial velocity planets will enter into a
transiting configuration, based upon assumptions regarding their
orbital inclinations. We finally compare periastron argument
uncertainties to the expected precession timescales and suggest
orbital refinement as a means to measure this effect.


\section{Transit Probability}
\label{tranprob}

Here we briefly describe the fundamentals of the geometric transit
probability for both circular and eccentric orbits. For a detailed
description we refer the reader to \citet{kan08}.

In the case of a circular orbit, the geometric transit probability is
defined as follows
\begin{equation}
  P_t = \frac{R_p + R_\star}{a}
\end{equation}
where $a$ is the semi-major axis and $R_p$ and $R_\star$ are the radii
of the planet and host star respectively. More generally, both the
transit and eclipse probabilities are inversely proportional to the
star--planet separation where the planet passes the star-observer
plane that is perpendicular to the plane of the planetary orbit. The
star--planet separation as a function of orbital eccentricity $e$ is
given by
\begin{equation}
  r = \frac{a (1 - e^2)}{1 + e \cos f}.
  \label{separation}
\end{equation}
where $f$ is the true anomaly, which describes the location of the
planet in its orbit, and so is a time dependent variable as the planet
orbits the star. For a transit event to occur the condition of $\omega
+ f = \pi / 2$ must be fulfilled \citep{kan07}, where $\omega$ is the
argument of periastron, and so we evaluate the above equations with
this condition in place. The geometric transit probability may thus be
re-expressed as
\begin{equation}
  P_t = \frac{(R_p + R_\star)(1 + e \cos (\pi/2 - \omega))}{a (1 -
    e^2)}
  \label{tranprobeqn}
\end{equation}
which is valid for any orbital eccentricity. Note that these equations
are independent of the true inclination of the planet's orbital plane.

\begin{figure}
  \includegraphics[angle=270,width=8.2cm]{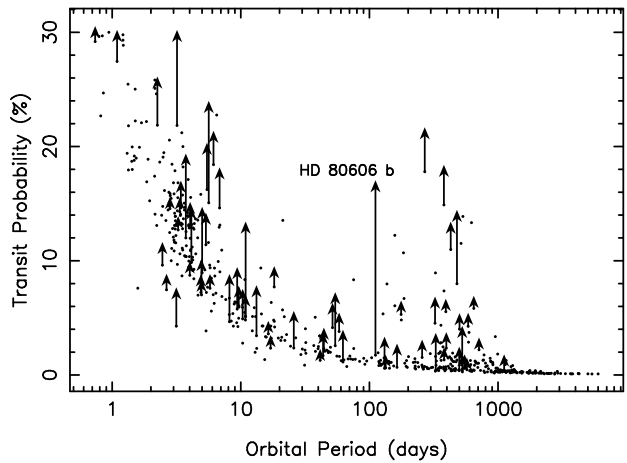}
  \caption{Transit probability for a sample of the known exoplanets as
    a function of orbital period. In cases where a change in $\omega$
    from current to $90\degr$ results in a transit probability
    improvement $> 1$\%, a vertical arrow indicates the improvement.}
  \label{tranprobfig}
\end{figure}

Given the sensitivity of transit probability to the argument of
periastron, it is useful to assess how the probabilities for the known
exoplanets would alter if their orientation was that most favorable
for transit detection: $\omega = 90\degr$. We extracted data from the
Exoplanet Data Explorer\footnote{\tt http://exoplanets.org/}
\citep{wri11} which include the orbital parameters and host star
properties for 592 planets and are current as of 30th June 2012. For
each planet, we calculate transit probabilities for two cases: (1)
using the current value of $\omega$, and (2) using $\omega =
90\degr$. The transit probabilities for case (1) are shown in Figure
\ref{tranprobfig}. Those planets whose case (2) probabilities are
improved by $> 1$\% are indicated by a vertical arrow to the improved
probability. There are several features of note in this figure.  The
relatively high transit probabilities between 100 and 1000 days are
due to giant host stars whose large radii dominates the probabilities
(see Equation \ref{tranprobeqn})). There are several cases of
substantially improved transit probability, most particularly
HD~80606~b, which is labelled in the figure. The following sections
investigate the periastron precession required to produce such an
increase in transit probability.


\section{Amplitude of Periastron (Apsidal) Precession}
\label{precession}

Periastron (or apsidal) precession is the gradual rotation of the
major axis which joins the orbital apsides within the orbital
plane. The result of this precession is that the argument of
periastron becomes a time dependent quantity. There are a variety of
factors which can lead to periastron precession, such as general
relativity (GR), stellar quadrupole moments, mutual star--planet tidal
deformations, and perturbations from other planets \citep{jor08}. For
Mercury, the perihelion precession rate due to general relativistic
effects is 43$\arcsec$/century (0.0119$\degr$/century). By comparison,
the precession due to perturbations from the other Solar System
planets is 532$\arcsec$/century (0.148$\degr$/century) while the
oblateness of the Sun (quadrupole moment) causes a negligible
contribution of 0.025$\arcsec$/century (0.000007$\degr$/century)
\citep{cle47,ior05}.

Here we adopt the formalism of \citet{jor08} in evaluating the
amplitude of the periastron precession. We first define the orbital
angular frequency as
\begin{equation}
  n \equiv \sqrt{ \frac{G M_\star}{a^3} } = \frac{2 \pi}{P}
  \label{angfreq}
\end{equation}
where $G$ is the gravitational constant, $M_\star$ is the mass of the
host star, and $P$ is the orbital period of the planet. The total
periastron precession is the sum of the individual effects as follows
\begin{equation}
  \dot{\omega}_{\mathrm{total}} = \dot{\omega}_{\mathrm{GR}} +
    \dot{\omega}_{\mathrm{quad}} + \dot{\omega}_{\mathrm{tide}} +
    \dot{\omega}_{\mathrm{pert}}
\end{equation}
where the precession components consist of the precession due to GR,
stellar quadrupole moment, tidal deformations, and planetary
perturbations respectively. \citet{jor08} conveniently express these
components in units of degrees per century. The components of
$\dot{\omega}_{\mathrm{quad}}$ and $\dot{\omega}_{\mathrm{tide}}$ have
$a^{-2}$ and $a^{-5}$ dependencies respectively. Since we are mostly
concerned with long-period planets in single-planet systems, we
consider here only the precession due to general relativity since this
is the dominant component in such cases. This imposes a lower limit on
the total precession of the system, particularly for multi-planet
systems. This precession is given by the following equation
\begin{equation}
  \dot{\omega}_{\mathrm{GR}} = \frac{7.78}{(1-e^2)} \left(
  \frac{M_\star}{M_\odot} \right) \left( \frac{a}{0.05 / \mathrm{AU}}
  \right)^{-1} \left( \frac{P}{\mathrm{day}} \right)^{-1}
  \label{grprec}
\end{equation}
with units in degrees per century.

\begin{figure}
  \includegraphics[angle=270,width=8.2cm]{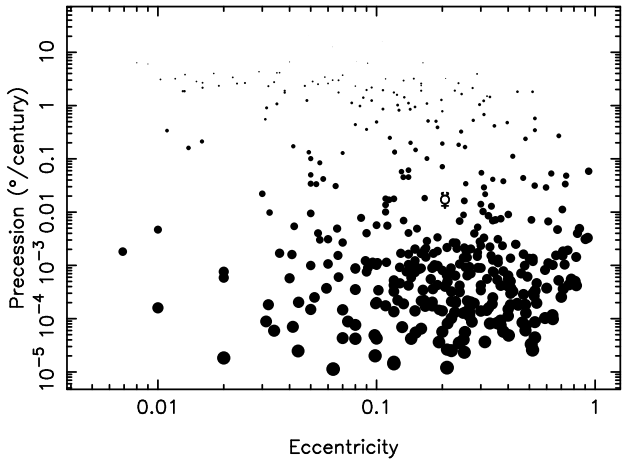}
  \caption{Calculated GR periastron precession rates plotted as a
    function of eccentricity for the known exoplanets with Keplerian
    orbital solutions. The radius of the points is logarithmically
    proportional to the orbital period of the planet. The symbol for
    Mercury is used to indicate its position on the plot.}
  \label{eccprec}
\end{figure}

To examine this precession effect for the known exoplanets, we use the
data extracted from the Exoplanet Data Explorer, described in Section
\ref{tranprob}. The GR precession rates for these planets are shown in
Figure \ref{eccprec} as a function of eccentricity, where the radius
of the point for each planet is logarithmically scaled with the
orbital period. As a Solar System example, the precession rate for
Mercury is shown using the appropriate symbol. There are two distinct
populations apparent in Figure \ref{eccprec} for which the divide
occurs at a periastron precession of $\sim 0.1\degr$/century. It is no
coincidence that this divide corresponds to the known relative dearth
of planets in the semi-major axis range of 0.1--0.6~AU
\citep{bur07b,cum08,cur09}.

\begin{figure}
  \includegraphics[angle=270,width=8.2cm]{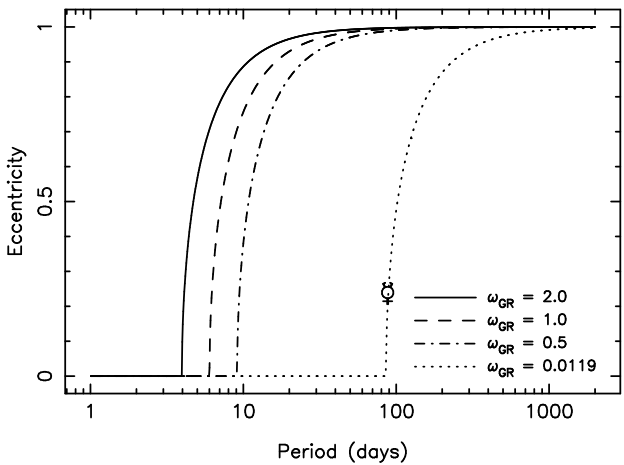}
  \caption{Lines of constant GR periastron precession as a function of
    orbital period and eccentricity, assuming a solar-mass host
    star. The eccentricity of the orbit only plays a significant role
    at very large values ($e > 0.8$). The symbol for Mercury is used
    to indicate its position on the plot.}
  \label{eccperiod}
\end{figure}

As expected from Equation \ref{grprec}, the amplitude of the
precession is dominated by the orbital period rather than the orbital
eccentricity. Thus, even planets in eccentric orbits do not exhibit
significant GR precession at longer periods. This is further
demonstrated in Figure \ref{eccperiod} where we show lines of constant
precession as a function of period and eccentricity for a solar-mass
host star. This shows that the GR periastron precession is almost
independent of orbital eccentricity except at extreme values of $e >
0.8$. Once again, the location of Mercury on the plot is indicated
using the appropriate symbol.

As noted by \citet{mir02} and \citet{jor08}, the total precession time
scales are large. Thus what really matters is the rate of change of
the periastron argument and quantifying when it is worth returning to
a particular target for re-investigation. This is the context of our
analysis in Section \ref{transit}.


\subsection{Nodal (Orbital Plane) Precession}

For completeness, we briefly consider the effects of nodal precession.
Nodal precession occurs when the orbital plane precesses around the
total angular momentum vector, which is usually aligned with the
rotation axis of the host star. The precession is caused by the
oblateness of the star which results in a non-zero gravitational
quadropole field. This has the potential to be the dominant source of
precession when the orbit is polar. For example, the nodal precession
for the near-polar retrograde orbit of WASP-33~b has been calculated
by \citet{ior11} to be $9 \times 10^9$ times larger than that induced
on the orbit of Mercury by the oblateness of the Sun.

A description of nodal precession and its effect on transit durations
has been provided by \citet{mir02}. The frequency of nodal precession
can be expressed as
\begin{equation}
  \Omega = n \frac{R_\star^2}{a^2} \frac{3 J_2}{4} \sin 2i
\end{equation}
where $n$ is the orbital angular frequency described in Equation
\ref{angfreq}, $J_2$ is the quadrupole moment, and $i$ is the orbital
inclination relative to the stellar equatorial plane. A typical
quadrupole moment for the star may be approximated as $J_2 \sim
10^{-6}$ and one may expect a relatively aligned orbit such that $\sin
2i \sim 0.1$. For a typical hot Jupiter, values for $a$ are $10
R_\star$, whereas for Mercury $a = 83 R_\star$. Since the nodal
precession is in units of the orbital angular frequency, one can see
that the resulting precession rate is typically several orders of
magnitude smaller than that of a hot Jupiter, even at the orbital
distance of Mercury. This effect is generally only considered for
circular orbits, most notably for short-period orbits that are the
most frequently encountered nature of known transiting planets. Here,
we are considering longer period eccentric orbits where this is a much
smaller effect on the orbital dynamics of the planet.


\section{Cyclic Transit Effects}
\label{transit}

As discussed in Section \ref{tranprob}, the transit probability for a
given planet is a function of the periastron argument for orbits with
non-zero eccentricity \citep{kan08}. The precession of the periastron
argument thus leads to a cyclic change in the transit
probability. Here we quantify this cyclic behaviour and determine
rates of change and total timescales.

\begin{figure*}
  \begin{center}
    \includegraphics[angle=270,width=15.0cm]{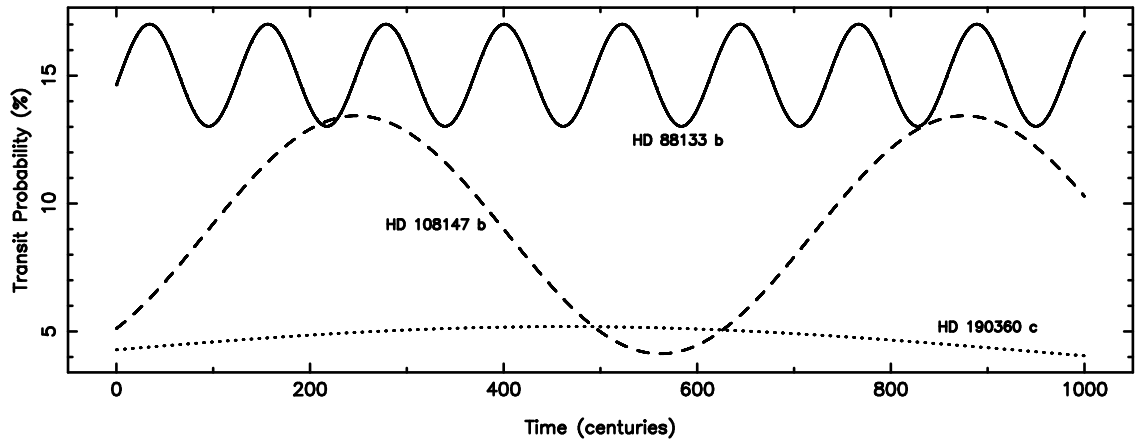}
  \end{center}
  \caption{Cyclic transit probabilities resulting from GR periastron
    precession for three known exoplanets: HD~88133~b, HD~108147~b,
    and HD~190360~c. This is shown from the present epoch and
    projected 100,000 years from now.}
  \label{cyclicplot}
\end{figure*}

Using the periastron precession rates calculated in Section
\ref{precession} and combining these with the transit probability
equations of Section \ref{tranprob} allows us to compute the time
dependent transit probability for each planet. Recall also that this
cyclic behaviour will only occur for planets which have non-zero
eccentricities. Shown in Figure \ref{cyclicplot} are three examples of
this time dependence over a period of 100,000 years. When viewing such
a plot one is tempted to interpret the cyclic variability in terms of
the orbital period, however this variation is caused by the periastron
precession, not the orbital period. There is, of course, some period
dependency involved, in that shorter period orbits will tend to have a
higher cyclic frequency. The planets shown here (HD~88133~b,
HD~108147~b, and HD~190360~c) have orbital periods of 3.4, 10.9, and
17.1 days respectively \citep{but06,wri09}. HD~108147~b, in
particular, displays very large amplitude variations due to the
relatively high eccentricity of its orbit ($e = 0.53$). HD~190360~c
has a smaller eccentricity and periastron precession rate, which leads
to a cyclic timescale much greater than 100,000 years.

We have performed these calculations for a subset of the known
exoplanets using the data extracted from the Exoplanet Data Explorer,
described in Section \ref{tranprob}. We restrict our sample to those
planets which are not known to transit and have non-zero
eccentricities. The results of these calculations are shown in Table
\ref{cyclictable} for 60 of the planets. The calculated values include
the periastron precession rate ($\dot{\omega}_{\mathrm{GR}}$), transit
probability ($P_t$), maximum transit probability at $\omega = 90\degr$
($P_t'$), time from the current epoch until maximum transit
probability ($\Delta t$), and the transit probability rate of change
($dP_t/dt$). The table has been sorted according to $dP_t/dt$ which is
presented in units of \%/century. The $dP_t/dt$ values have been
calculated from the current epoch over the coming century and thus
represents the present rate of change. The importance of this is that
$dP_t/dt$ is not constant and indeed can have negative values as the
periastron argument rotates past $\omega = 90\degr$. Specifically,
$dP_t/dt$ will be negative for $90\degr < \omega < 270\degr$ and
positive elsewhere. This further restricts the planets considered to
those whose current $\omega$ falls in this range such that $dP_t/dt >
0$.

It can be clearly seen that the time required to reach maximum transit
probability is immense, certainly beyond the lifetime of anyone
reading this work. However, the rate of change can yield an improved
idea of which planets may have a measurable change in configuration.
Consider the case of HD~156846~b, whose orbital parameters and transit
potential have been studied in detail by \citet{kan11}. This is one of
the planets in the table with the longest period and also has one of
the highest orbital eccentricities. The transit probability is
relatively high for this planet and is close to the maximum
probability since $\omega$ only needs to change by $38\degr$. Even so,
observations of the periastron precession are unlikely for the
timescales involved. By contrast, the hot Saturn HD~88133~b discovered
by \citet{fis05} has the highest transit probability rate of change.

\begin{deluxetable*}{lcccccccc}
  \tablecolumns{9}
  \tablewidth{0pc}
  \tablecaption{\label{cyclictable} Exoplanet Periastron Precession,
    Transit Probabilities, and Timescales}
  \tablehead{
    \colhead{Planet} &
    \colhead{$P$ (days)} &
    \colhead{$e$} &
    \colhead{$\omega$} &
    \colhead{$\dot{\omega}_{\mathrm{GR}}$ ($\degr$/cent)} &
    \colhead{$P_t$ (\%)} &
    \colhead{$P_t'$ (\%)$^1$} &
    \colhead{$\Delta t$ (cent)$^2$} &
    \colhead{$dP_t/dt$ (\%/cent)$^3$}
  }
  \startdata
HD 88133 b    &     3.42 & 0.13 & 349.0 & 2.9490 & 14.6 & 17.0 &       34.2 &  0.101368 \\
HD 76700 b    &     3.97 & 0.09 &  30.0 & 2.1838 & 12.9 & 13.5 &       27.5 &  0.038099 \\
HD 73256 b    &     2.55 & 0.03 & 337.3 & 4.3194 & 16.1 & 16.8 &       26.1 &  0.033421 \\
HD 108147 b   &    10.90 & 0.53 & 308.0 & 0.5732 &  5.1 & 13.4 &      247.7 &  0.028841 \\
HD 102956 b   &     6.49 & 0.05 &  12.0 & 1.2451 & 22.8 & 23.6 &       62.6 &  0.022932 \\
BD -08 2823 b &     5.60 & 0.15 &  30.0 & 0.9420 & 11.6 & 12.4 &       63.7 &  0.022925 \\
HD 7924 b     &     5.40 & 0.17 &  25.0 & 1.0901 &  7.2 &  7.9 &       59.6 &  0.019663 \\
HD 68988 b    &     6.28 & 0.12 &  31.4 & 1.0214 &  8.7 &  9.1 &       57.4 &  0.015372 \\
HD 1461 b     &     5.77 & 0.14 &  58.0 & 1.1102 &  9.4 &  9.6 &       28.8 &  0.011871 \\
HD 217107 b   &     7.13 & 0.13 &  24.4 & 0.8192 &  6.9 &  7.4 &       80.1 &  0.010737 \\
HD 168746 b   &     6.40 & 0.11 &  17.0 & 0.8587 &  7.2 &  7.7 &       85.0 &  0.010620 \\
HD 149143 b   &     4.07 & 0.02 &   0.0 & 2.1614 & 15.5 & 15.8 &       41.6 &  0.009380 \\
HD 162020 b   &     8.43 & 0.28 &  28.4 & 0.5283 &  4.7 &  5.3 &      116.6 &  0.009352 \\
HD 187123 b   &     3.10 & 0.01 &  24.5 & 3.0953 & 13.5 & 13.6 &       21.1 &  0.006702 \\
HD 47186 b    &     4.08 & 0.04 &  59.0 & 1.8945 & 11.0 & 11.0 &       16.4 &  0.006690 \\
BD -10 3166 b &     3.49 & 0.02 & 334.0 & 2.3445 &  8.7 &  8.9 &       49.5 &  0.006174 \\
HD 69830 b    &     8.67 & 0.10 & 340.0 & 0.4923 &  5.4 &  6.1 &      223.5 &  0.004500 \\
HD 190360 c   &    17.11 & 0.24 &   5.2 & 0.1833 &  4.3 &  5.2 &      462.8 &  0.003171 \\
upsilon And b &     4.62 & 0.01 &  51.0 & 1.8588 & 12.0 & 12.0 &       21.0 &  0.003147 \\
HD 179079 b   &    14.48 & 0.12 & 357.0 & 0.2481 &  5.3 &  6.0 &      374.8 &  0.002675 \\
51 Peg b      &     4.23 & 0.01 &  58.0 & 1.8602 & 10.1 & 10.1 &       17.2 &  0.002169 \\
HD 10180 c    &     5.76 & 0.08 & 279.0 & 1.1232 &  7.6 &  8.8 &      152.2 &  0.002059 \\
HIP 57274 b   &     8.14 & 0.19 &  81.0 & 0.5075 &  5.3 &  5.3 &       17.7 &  0.001124 \\
HD 147018 b   &    44.24 & 0.47 & 336.0 & 0.0437 &  2.3 &  4.1 &     2607.5 &  0.000922 \\
HD 16417 b    &    17.24 & 0.20 &  77.0 & 0.1935 &  6.3 &  6.4 &       67.2 &  0.000801 \\
HD 10180 d    &    16.36 & 0.14 & 292.0 & 0.2001 &  3.6 &  4.7 &      789.4 &  0.000780 \\
HD 163607 b   &    75.29 & 0.73 &  78.7 & 0.0336 &  8.3 &  8.4 &      336.7 &  0.000406 \\
HD 224693 b   &    26.73 & 0.05 &   6.0 & 0.1008 &  3.2 &  3.4 &      833.1 &  0.000283 \\
4 UMa b       &   269.30 & 0.43 &  23.8 & 0.0025 & 17.8 & 21.7 &    26488.3 &  0.000263 \\
61 Vir c      &    38.02 & 0.14 & 341.0 & 0.0453 &  2.1 &  2.5 &     2405.3 &  0.000227 \\
HD 102117 b   &    20.81 & 0.12 & 279.0 & 0.1351 &  3.3 &  4.2 &     1266.0 &  0.000169 \\
HD 43691 b    &    36.96 & 0.14 & 290.0 & 0.0612 &  2.6 &  3.4 &     2612.5 &  0.000152 \\
70 Vir b      &   116.69 & 0.40 & 358.7 & 0.0090 &  1.9 &  2.7 &    10097.0 &  0.000124 \\
HD 156846 b   &   359.51 & 0.85 &  52.2 & 0.0049 &  4.4 &  4.8 &     7699.5 &  0.000116 \\
HD 16141 b    &    75.52 & 0.25 &  42.0 & 0.0163 &  2.3 &  2.5 &     2950.5 &  0.000105 \\
GJ 785 b      &    74.39 & 0.30 &  15.0 & 0.0141 &  1.4 &  1.6 &     5332.1 &  0.000089 \\
HIP 57274 c   &    32.03 & 0.05 & 356.2 & 0.0500 &  1.9 &  2.0 &     1876.0 &  0.000083 \\
HD 4113 b     &   526.62 & 0.90 & 317.7 & 0.0031 &  0.9 &  4.3 &    42500.7 &  0.000082 \\
rho CrB b     &    39.84 & 0.06 & 303.0 & 0.0419 &  2.3 &  2.6 &     3510.1 &  0.000055 \\
HD 45652 b    &    43.60 & 0.38 & 273.0 & 0.0380 &  1.7 &  3.8 &     4660.2 &  0.000036 \\
HD 20868 b    &   380.85 & 0.75 & 356.2 & 0.0019 &  1.3 &  2.4 &    48796.3 &  0.000035 \\
61 Vir d      &   123.01 & 0.35 & 314.0 & 0.0072 &  0.8 &  1.5 &    19010.2 &  0.000034 \\
55 Cnc c      &    44.38 & 0.05 &  57.4 & 0.0335 &  2.1 &  2.1 &      972.3 &  0.000033 \\
HD 60532 b    &   201.30 & 0.28 & 351.9 & 0.0040 &  1.6 &  2.1 &    24667.9 &  0.000032 \\
HD 145457 b   &   176.30 & 0.11 & 300.0 & 0.0056 &  5.3 &  6.5 &    26940.2 &  0.000032 \\
GJ 581 d      &    66.64 & 0.25 & 356.0 & 0.0089 &  0.7 &  0.9 &    10602.7 &  0.000029 \\
HD 5891 b     &   177.11 & 0.07 & 351.0 & 0.0049 &  4.8 &  5.2 &    20191.3 &  0.000027 \\
HD 1237 b     &   133.71 & 0.51 & 290.7 & 0.0072 &  0.6 &  1.8 &    22236.9 &  0.000027 \\
HD 17092 b    &   359.90 & 0.17 & 347.4 & 0.0020 &  2.8 &  3.4 &    52495.0 &  0.000016 \\
HD 22781 b    &   528.07 & 0.82 & 315.9 & 0.0014 &  0.4 &  1.8 &    93315.9 &  0.000015 \\
HD 107148 b   &    48.06 & 0.05 &  75.0 & 0.0342 &  2.1 &  2.1 &      439.0 &  0.000015 \\
BD +48 738 b  &   392.60 & 0.20 & 358.9 & 0.0008 &  5.5 &  6.7 &   113252.5 &  0.000015 \\
HD 180314 b   &   396.03 & 0.26 & 303.1 & 0.0019 &  2.3 &  3.8 &    78180.5 &  0.000014 \\
HIP 14810 c   &   147.77 & 0.15 & 327.3 & 0.0049 &  1.1 &  1.4 &    25078.6 &  0.000013 \\
HD 8574 b     &   227.00 & 0.30 &  26.6 & 0.0028 &  1.1 &  1.3 &    22763.8 &  0.000013 \\
HD 216770 b   &   118.45 & 0.37 & 281.0 & 0.0075 &  0.8 &  1.8 &    22495.8 &  0.000012 \\
HD 93083 b    &   143.58 & 0.14 & 333.5 & 0.0041 &  1.1 &  1.3 &    28700.1 &  0.000010 \\
HD 11977 b    &   711.00 & 0.40 & 351.5 & 0.0006 &  2.2 &  3.3 &   153321.8 &  0.000010 \\
HD 222582 b   &   572.38 & 0.73 & 319.0 & 0.0010 &  0.5 &  1.5 &   126637.2 &  0.000009 \\
HD 231701 b   &   141.60 & 0.10 &  46.0 & 0.0057 &  1.2 &  1.2 &     7727.6 &  0.000008 \\
  \enddata
  \tablenotetext{$^1$}{$P_t'$ refers to the transit probability where
    $\omega = 90\degr$.}
  \tablenotetext{$^2$}{$\Delta t$ refers to the time until $P_t'$
    occurs.}
  \tablenotetext{$^3$}{$dP_t/dt$ is calculated over the coming century
    but is a time dependent quantity.}
\end{deluxetable*}


\section{Conclusions}

Transiting planets have become an essential component of exoplanetary
science due to the exceptional opportunities they present for
characterization of these planets. Many of the known exoplanets
discovered through the radial velocity technique are currently not
known to transit. However, transit probabilities can be substantially
improved if the periastron argument approaches $\omega =
90\degr$. Since, for eccentric orbits, the periastron argument is time
dependent as a result of their precession, planets which do not
transit at the present epoch may transit in the future and vice
versa. The planet Mercury falls quite central to the current
distribution of calculated periastron precessions for the known
exoplanets. This distribution has an eccentricity dependence but is
most strongly affected by the orbital period. If a precession rate for
a given planet is found to be markedly different from our calculations
then this could be indicative of further, as yet undiscovered planets
in that system. These additional planets would normally be detected
from the radial velocity data unless insufficient observations allow
them to remain hidden.

The periastron precession leads to a cyclic transit probability
variation for all exoplanets with non-zero eccentricities. Timescales
vary enormously but will likely lead to many of these planets
transiting their host stars at some point in the future.  A reasonable
question to ask at this point is if the periastron arguments of the
known planets are known with sufficient precision to detect precession
in any acceptable timeframe. Once again, we exploit the data extracted
from the Exoplanet Data Explorer, described in Section
\ref{tranprob}. The uncertainties associated with the values of
$\omega$ for all these planets have a mean of $28\degr$ and a median
of $15\degr$. This is much higher than the precession effects shown in
Table \ref{cyclictable}. A program of refining the orbits of the known
exoplanets, such as that described by \citet{kan09a}, would result in
many of these precession effects to be detectable in reasonable time
frames. For example the first planet in the table, HD~88133~b, has a
precession rate that will cause a shift of $\sim 0.3\degr$ per
decade. Uncertainties on $\omega$ of less than one degree are not
unsual and can certainly be achieved for those planets in particularly
eccentric orbits. The exoplanet HD~156846~b has a current $\omega$
uncertainty of $0.16\degr$ \citep{kan11} which demonstrates that such
refinement is possible even for relatively long-period planets. More
data and longer time baselines will produce subsequent improvements
for many more planets which can result in the detection of the
precession for high-precession cases.

The relevance of this work may be extended to the Kepler mission which
has detected many candidate multi-planet systems
\citep{bor11a,bor11b,bat12}, most of which are likely to be real
exoplanets \citep{lis12}. Due to simply geometric transit
probabilities, most of these systems will certainly have planets which
are not transiting the host star at present. The known transiting
multi-planet systems are largely in circular orbits, but may have
periastron precession due to perturbations from other planets leading
to an eventual transit from currently non-transiting planets in the
system. For example, Kepler-19~c is known to exist from Transit Timing
Variations of the inner planet, but does not currently have a
detectable transit signature. Similarly, some of these planets will
cease exhibiting an observable transit signature. Issues such as
these are important for considering the completeness of these surveys
in determining multi-planetary system architectures.


\section*{Acknowledgements}

The authors would like to thank David Ciardi and Solange Ramirez for
several useful discussions and the numerous people who have requested
that we perform this study. We would also like to thank the anonymous
referee, whose comments greatly improved the quality of the paper. JH
gratefully acknowledges the financial support of the Australian
government through ARC Grant DP0774000.  This research has made use of
the Exoplanet Orbit Database and the Exoplanet Data Explorer at
exoplanets.org. This research has also made use of the NASA Exoplanet
Archive, which is operated by the California Institute of Technology,
under contract with the National Aeronautics and Space Administration
under the Exoplanet Exploration Program.



\begin{thebibliography}{}

\bibitem[\protect\citeauthoryear{Agol et al.}{2010}]{ago10} Agol, E.,
  Cowan, N.B., Knutson, H.A., Deming, D., Steffan, J.H., Henry, G.W.,
  Charbonneau, D. 2010, ApJ, 721, 1861
\bibitem[\protect\citeauthoryear{Barbieri et al.}{2007}]{bar07}
  Barbieri, M., et al. 2007, A\&A, 476, L13
\bibitem[\protect\citeauthoryear{Batalha et al.}{2012}]{bat12}
  Batalha, N.M., et al. 2012, ApJS, submitted (arXiv:1202.5852)
\bibitem[\protect\citeauthoryear{Borucki et al.}{2011a}]{bor11a}
  Borucki, W.J., et al. 2011, ApJ, 728, 117
\bibitem[\protect\citeauthoryear{Borucki et al.}{2011b}]{bor11b}
  Borucki, W.J., et al. 2011, ApJ, 736, 19
\bibitem[\protect\citeauthoryear{Burrows et al.}{2007}]{bur07a}
  Burrows, A., Hubeny, I., Budaj, J., Hubbard, W.B. 2007, ApJ, 661,
  502
\bibitem[\protect\citeauthoryear{Burkert \& Ida}{2007}]{bur07b} Burkert,
  A., Ida, S. 2007, ApJ, 660, 845
\bibitem[\protect\citeauthoryear{Butler et al.}{2006}]{but06}
  Butler, R.P. et al. 2006, ApJ, 646, 505
\bibitem[\protect\citeauthoryear{Carter \& Winn}{2010}]{car10}
  Carter, J.A., Winn, J.N. 2010, ApJ, 716, 850
\bibitem[\protect\citeauthoryear{Charbonneau et al.}{2000}]{cha00}
  Charbonneau, D., Brown, T.M., Latham, D.W., Mayor, M., 2000, ApJ,
  529, L45
\bibitem[\protect\citeauthoryear{Clemence}{1947}]{cle47} Clemence,
  G.M. 1947, RvMP, 19, 361
\bibitem[\protect\citeauthoryear{Cumming et al.}{2008}]{cum08}
  Cumming, A., Butler, R.P., Marcy, G.W., Vogt, S.S., Wright, J.T.,
  Fischer, D.A. 2008, PASP, 120, 531
\bibitem[\protect\citeauthoryear{Currie}{2009}]{cur09} Currie,
  T. 2009, ApJ, 694, L171
\bibitem[\protect\citeauthoryear{Damiani \& Lanza}{2011}]{dam11}
  Damiani, C., Lanza, A.F. 2011, A\&A, 535, 116
\bibitem[\protect\citeauthoryear{Deming et al.}{2007a}]{dem07} Deming,
  D., Richardson, L.J., Harrington, J. 2007, MNRAS, 378, 148
\bibitem[\protect\citeauthoryear{Fischer et al.}{2005}]{fis05}
  Fischer, D.A., et al. 2005, ApJ, 620, 481
\bibitem[\protect\citeauthoryear{Fortney et al.}{2007}]{for07}
  Fortney, J.J., Marley, M.S., Barnes, J.W., 2007, ApJ, 659, 1661
\bibitem[\protect\citeauthoryear{Henry et al.}{2000}]{hen00} Henry,
  G.W., Marcy, G.W., Butler, R.P., Vogt, S.S., 2000, ApJ, 529, L41
\bibitem[\protect\citeauthoryear{Heyl \& Gladman}{2007}]{hey07} Heyl,
  J.S., Gladman, B.J. 2007, MNRAS, 377, 1511
\bibitem[\protect\citeauthoryear{Iorio}{2005}]{ior05} Iorio, L. 2005,
  A\&A, 433, 385
\bibitem[\protect\citeauthoryear{Iorio}{2011}]{ior11} Iorio, L. 2011,
  Ap\&SS, 331, 485
\bibitem[\protect\citeauthoryear{Jord\'an \& Bakos}{2008}]{jor08}
  Jord\'an, A., Bakos, G.A. 2008, ApJ, 685, 543
\bibitem[\protect\citeauthoryear{Kane}{2007}]{kan07} Kane, S.R.
  2007, MNRAS, 380, 1488
\bibitem[\protect\citeauthoryear{Kane \& von Braun}{2008}]{kan08}
  Kane, S.R., von Braun, K. 2008, ApJ, 689, 492
\bibitem[\protect\citeauthoryear{Kane \& von Braun}{2009}]{kan09b}
  Kane, S.R., von Braun, K., 2009, PASP, 121, 1096
\bibitem[\protect\citeauthoryear{Kane et al.}{2009}]{kan09a} Kane,
  S.R., Mahadevan, S., von Braun, K., Laughlin, G., Ciardi, D.R.
  2009, PASP, 121, 1386
\bibitem[\protect\citeauthoryear{Kane et al.}{2011}]{kan11} Kane,
  S.R., et al. 2011, ApJ, 733, 28
\bibitem[\protect\citeauthoryear{Knutson et al.}{2009a}]{knu09a}
  Knutson, H.A., et al. 2009a, ApJ, 690, 822
\bibitem[\protect\citeauthoryear{Knutson et al.}{2009b}]{knu09b}
  Knutson, H.A., Charbonneau, D., Cowan, N.B., Fortney, J.J., Showman,
  A.P., Agol, E., Henry, G.W. 2009b, ApJ, 703, 769
\bibitem[\protect\citeauthoryear{Laughlin et al.}{2009}]{lau09}
  Laughlin, G., Deming, D., Langton, J., Kasen, D., Vogt, S., Butler,
  P., Rivera, E., Meschiari, S. 2009, Nature, 457, 562
\bibitem[\protect\citeauthoryear{Lissauer et al.}{2012}]{lis12}
  Lissauer, J.J., et al. 2012, ApJ, 750, 112
\bibitem[\protect\citeauthoryear{Miralda-Escud\'e}{2002}]{mir02}
  Miralda-Escud\'e, J. 2002, ApJ, 564, 1019
\bibitem[\protect\citeauthoryear{P\'al \& Kocsis}{2008}]{pal08}
  P\'al, A., Kocsis, B. 2008, MNRAS, 389, 191
\bibitem[\protect\citeauthoryear{Ragozzine \& Wolf}{2009}]{rag09}
  Ragozzine, D., Wolf, A.S. 2009, ApJ, 698, 1778
\bibitem[\protect\citeauthoryear{Seager et al.}{2007}]{sea07} Seager,
  S., Kuchner, M., Hier-Majumder, C.A., Militzer, B. 2007, ApJ, 669,
  1279
\bibitem[\protect\citeauthoryear{Wright et al.}{2009}]{wri09} Wright,
  J.T., Upadhyay, S., Marcy, G.W., Fischer, D.A., Ford, E.B., Johnson,
  J.A. 2009, ApJ, 693, 1084
\bibitem[\protect\citeauthoryear{Wright et al.}{2011}]{wri11}
  Wright, J.T., et al., 2011, PASP, 123, 412

\end{thebibliography}
\end{document}